\documentclass[%
 reprint,
superscriptaddress,
 amsmath,amssymb,
 aps,
]{revtex4-2}

\usepackage{graphicx}
\usepackage{dcolumn}
\usepackage{bm}

\begin{document}
\newcommand{\argmax}{\mathop{\rm arg~max}\limits}
\newcommand{\argmin}{\mathop{\rm arg~min}\limits}

\preprint{APS/123-QED}

\title{Superluminal Motion-Assisted 4-Dimensional Light-in-Flight Imaging}

\author{Kazuhiro Morimoto} \thanks{These authors contributed equally} \affiliation{Advanced Quantum Architecture Laboratory, Ecole polytechnique fédérale de Lausanne, Neuchâtel 2002, Switzerland}  \affiliation{Device Research \& Design Department, Canon Inc., Kanagawa 212-8602, Japan}
\author{Ming-Lo Wu} \thanks{These authors contributed equally} \affiliation{Advanced Quantum Architecture Laboratory, Ecole polytechnique fédérale de Lausanne, Neuchâtel 2002, Switzerland}
\author{Andrei Ardelean} \thanks{These authors contributed equally} \affiliation{Advanced Quantum Architecture Laboratory, Ecole polytechnique fédérale de Lausanne, Neuchâtel 2002, Switzerland}
\author{Edoardo Charbon} \email{edoardo.charbon@epfl.ch} \affiliation{Advanced Quantum Architecture Laboratory, Ecole polytechnique fédérale de Lausanne, Neuchâtel 2002, Switzerland}

\date{\today}

\begin{abstract}
Advances in high speed imaging techniques have opened new possibilities for capturing ultrafast phenomena such as light propagation in air or through media. Capturing light-in-flight in 3-dimensional $xyt$-space has been reported based on various types of imaging systems, whereas reconstruction of light-in-flight information in the fourth dimension $z$ has been a challenge. We demonstrate the first 4-dimensional light-in-flight imaging based on the observation of a superluminal motion captured by a new time-gated megapixel single-photon avalanche diode camera. A high resolution light-in-flight video is generated with no laser scanning, camera translation, interpolation, nor dark noise subtraction. A machine learning technique is applied to analyze the measured spatio-temporal data set. A theoretical formula is introduced to perform least-square regression, and extra-dimensional information is recovered without prior knowledge. The algorithm relies on the mathematical formulation equivalent to the superluminal motion in astrophysics, which is scaled by a factor of a quadrillionth. The reconstructed light-in-flight trajectory shows a good agreement with the actual geometry of the light path. Our approach could potentially provide novel functionalities to high speed imaging applications such as non-line-of-sight imaging and time-resolved optical tomography.\\
\\
\end{abstract}

\maketitle


\section{\label{sec:level1}INTRODUCTION}
Progress in high speed imaging techniques has enabled observation and recording of light propagation dynamics in free space as well as in transparent, translucent and scattering media. Various approaches for capturing the light-in-flight have been demonstrated: holographic techniques~\cite{Abramson,Abramson2,Abramson3}, photonic mixers~\cite{Heide}, time-encoded amplified imaging~\cite{Goda}, streak cameras~\cite{Velten,Velten2}, intensified CCDs~\cite{Faccio} and silicon CMOS image sensors~\cite{Etoh}. In recent years, 1- and 2-dimensional arrays of single-photon avalanche diode (SPAD) have been adopted for the light-in-flight imaging systems~\cite{Gariepy,Laurenzis,Warburton,OToole,Wilson,Sun}. These sensors boost data acquisition speed by employing pixel-parallel detection of time stamping with picosecond time resolution and single-photon sensitivity. However, the sensors are capable of sampling only 3-dimensional spatio-temporal information ($x$, $y$, $t$) of the target light, and hence the tracking of light-in-flight breaking outside a $xy$-plane is a challenge.\\
A method to reconstruct an extra-dimensional position information, $z$, from measured spatio-temporal information has been recently investigated by researchers~\cite{Laurenzis2}. The authors remark that different propagation angle with respect to $xy$-plane results in the different apparent velocity of light. Thus, comparing the measured spatio-temporal data set with this theory could give an estimation of $z$-component in the light propagation vector. Yet, the analysis is limited to a simplified case with single straight light paths within $xz$-plane passing a fixed point, and complete 4-dimensional reconstruction of light-in-flight in arbitrary paths remains to be verified. In addition, spatial resolution of the detector array is critical for accurate reconstruction of the 4-dimensional trace; the reported SPAD sensor resolutions, e.g. 64$\times$32~\cite{Sun}, 32$\times$32~\cite{Gariepy} and 256$\times$1~\cite{OToole} pixels, are not sufficient, whereas significant improvement of the estimation error for ($x$, $y$, $z$, $t$) is expected with orders of magnitude larger arrays, typically in the megapixels. CMOS sensors routinely achieve megapixel resolutions, e.g. 0.3 Mpixel resolution is reported for high speed imaging~\cite{Etoh}, however the necessary timing resolution is not reached and it is limited to 10 nanoseconds.\\
In this paper we demonstrate the 4-dimensional light-in-flight imaging based on the first time-gated megapixel SPAD camera~\cite{Morimoto}. In contrast to conventional time-correlated single-photon counting (TCSPC) approaches, where power- and area-consuming time-to-digital converter (TDC) circuits restrict the scaling of the pixel array~\cite{Veerappan,Carimatto,Henderson}, our time-gating approach~\cite{Morimoto} achieves a much more compact pixel circuit, suitable for large-scale arrays. Our camera achieves a megapixel format, while, at the same time ensuring a time resolution comparable to that of a TDC. Owing to an unprecedented high spatio-temporal resolution, a superluminal motion of laser pulse was successfully captured in a frame sequence. We then introduced a theoretical equation to fit the measured data set in 3-dimensional space. A 4-dimensional point cloud is computationally reproduced without prior knowledge, exhibiting a good agreement with the actual light propagation vectors.\\

\section{\label{sec:level2}PRINCIPLE AND EXPERIMENTAL SETUP}
\begin{figure}[b]
\includegraphics[width=1\columnwidth]{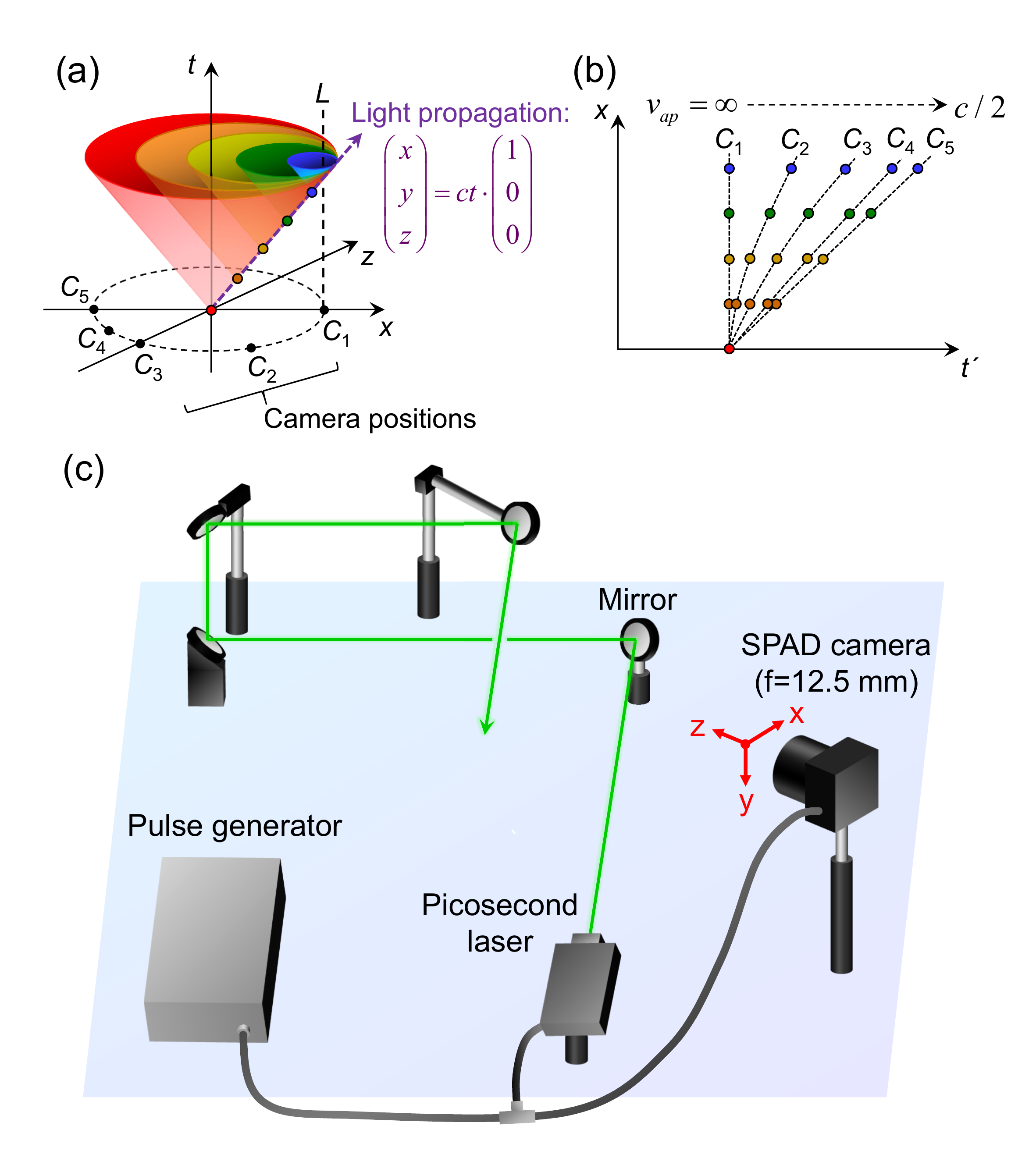}
\caption{\label{setup}Principle and experimental setup for 4-dimensional light-in-flight imaging. (a) Schematic illustration of light propagation and scattering in Minkowski space at $y$ = 0. Propagating light is depicted as a purple dashed arrow, and scattered light is shown as a light cone. (b) Schematic plot of $t^{\prime}$ dependence of $x$. The slope of the dashed line indicates the apparent velocity. (c) Experimental setup for light-in-flight imaging. Picosecond laser and SPAD camera are both controlled by a pulse generator. When performing the light-in-flight measurements, all the mirrors are confined by a large transparent acrylic box (not shown), and a fog generator forms a small amount of mist in the box to enhance the scattering of the laser in air. The origin of the Cartesian coordinate is set at the optical center of the lens of the SPAD camera.}
\end{figure}
\begin{figure*}[tb]
\includegraphics[width=1.7\columnwidth]{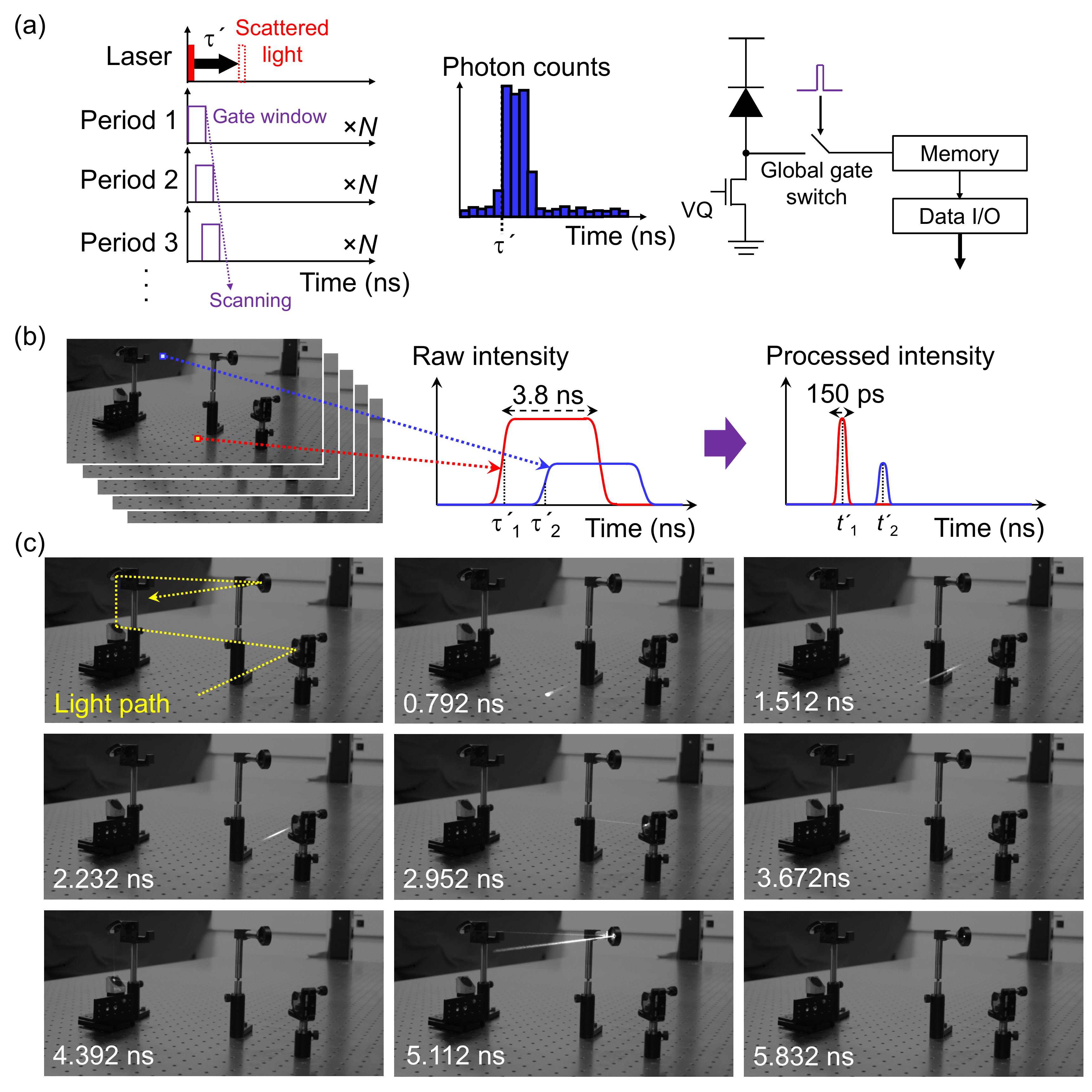}
\caption{\label{video_frames}Video generation of 3-dimensional light-in-flight imaging. (a) Principle of the time gate scanning method for acquiring time-of-arrival $\tau^{\prime}$ of impinging photons. (b) Data processing flow for generation of light-in-flight video. (c) The laser pulse propagates over time along the dashed line in the top-left image. A superluminal motion of laser pulse was observed in the latter frames. The light-in-flight video is taken under dark conditions, and each frame is superimposed with a 12-bit 2-dimensional intensity image independently taken with the same SPAD camera under room light. The image resolution is 1024$\times$500 pixels. }
\end{figure*}
To illustrate our method, we introduce a Minkowski space that is describing the spatio-temporal propagation of a laser pulse and scattered photons towards fixed observation points. Fig.~\ref{setup}(a) shows a schematic illustration of light-in-flight observation in the Minkowski space at $y$ = 0. A straight light propagation along $x$-axis is depicted as ($x$, $y$, $z$) = ($ct$, 0, 0), where $c$ is the speed of light. In the Minkowski space, this equation corresponds to a single tilted line crossing the origin (purple arrow in Figure~\ref{setup}(a)). A camera is located at the position $C_i$ ($i$ = 1,2,…,5). The observation of the light-in-flight at the camera is mediated by scattered light at each point on the path (colored dots on the purple arrow). The propagation of the scattered light from the points are described as light cones. Here we define t as the time at which the light propagating from a laser source reaches a point ($x$, $y$, $z$), called scattering point on the light path, and $t^{\prime}$ as a time at which the light scattered by the medium at ($x$, $y$, $z$) reaches the camera. Due to the finite speed of light, a time difference from $t$ to $t^{\prime}$ is proportional to the spatial distance between the scattering point and the camera location. In Minkowski space, the observation time $t^{\prime}$ can be visualized as an intersection point of the light cone and a vertical line $L$ passing $C_i$ (dotted line). The relation between the light position $x$ and the corresponding observation time $t^{\prime}$ is shown in Figure~\ref{setup}(b). The behavior is highly dependent on the camera position with respect to the scattering point, i.e. at certain points it appears to be faster than in others. For example, at $C_1$ where the light is coming towards the camera, the scattering light from all the points on the light path reaches the camera at the same time. The corresponding apparent velocity, defined as $v_{ap}$ = $dx/dt^{\prime}$, is infinite, or $t^{\prime}$ appears not to change irrespective of the distance of the scattering point ($x$, $y$, $z$) from the camera. At $C_5$, in contrast, the light propagating away from the camera is observed with the apparent velocity of $c$/2. For other scattering points corresponding to relative camera positions $C_2$, $C_3$, and $C_4$, one observes intermediate apparent velocities, depending on the relative position and angle of the line of sight between scattering point and camera. Hence, analyzing the apparent velocity provides an estimated light propagation vector with extra-dimensional information. The discussion can be generalized for higher dimensions; the detailed analysis of measured ($x$, $y$, $t$) data set and its local apparent velocity would enable a 4-dimensional reconstruction of the light-in-flight history. A 2D-projected intuitive visualization for the propagation of scattered light is provided in Supplementary Video 1. Note that angle-dependent apparent velocity of light can be seen in nature, for instance in an astronomical phenomenon called superluminal motion. In this case, relativistic jets of matter are emitted from radio galaxies, quasars, and other celestial objects, appearing to travel faster than light to the observer in a time scale of months to years. Light echoes are also known to produce superluminal motion~\cite{Bond,Mooley}.\\
Fig.~\ref{setup}(c) is the experimental setup to verify the 4-dimensional light-in-flight imaging. A 510 nm-laser (average power: 2 mW, frequency: 40 MHz, optical pulse width: 130 ps, PicoQuant GmbH, Berlin, Germany) and the megapixel SPAD camera are synchronized by a pulse generator. The emitted laser pulses are reflected by four mirrors to construct a 3-dimensional trajectory. The megapixel SPAD camera can be operated in intensity imaging mode to capture the background scene and in time gating mode to capture laser propagation~\cite{Morimoto}. 

\section{\label{sec:level3}3-DIMENSIONAL LIGHT-IN-FLIGHT IMAGING}
\begin{figure*}[tb]
\includegraphics[width=1.8\columnwidth]{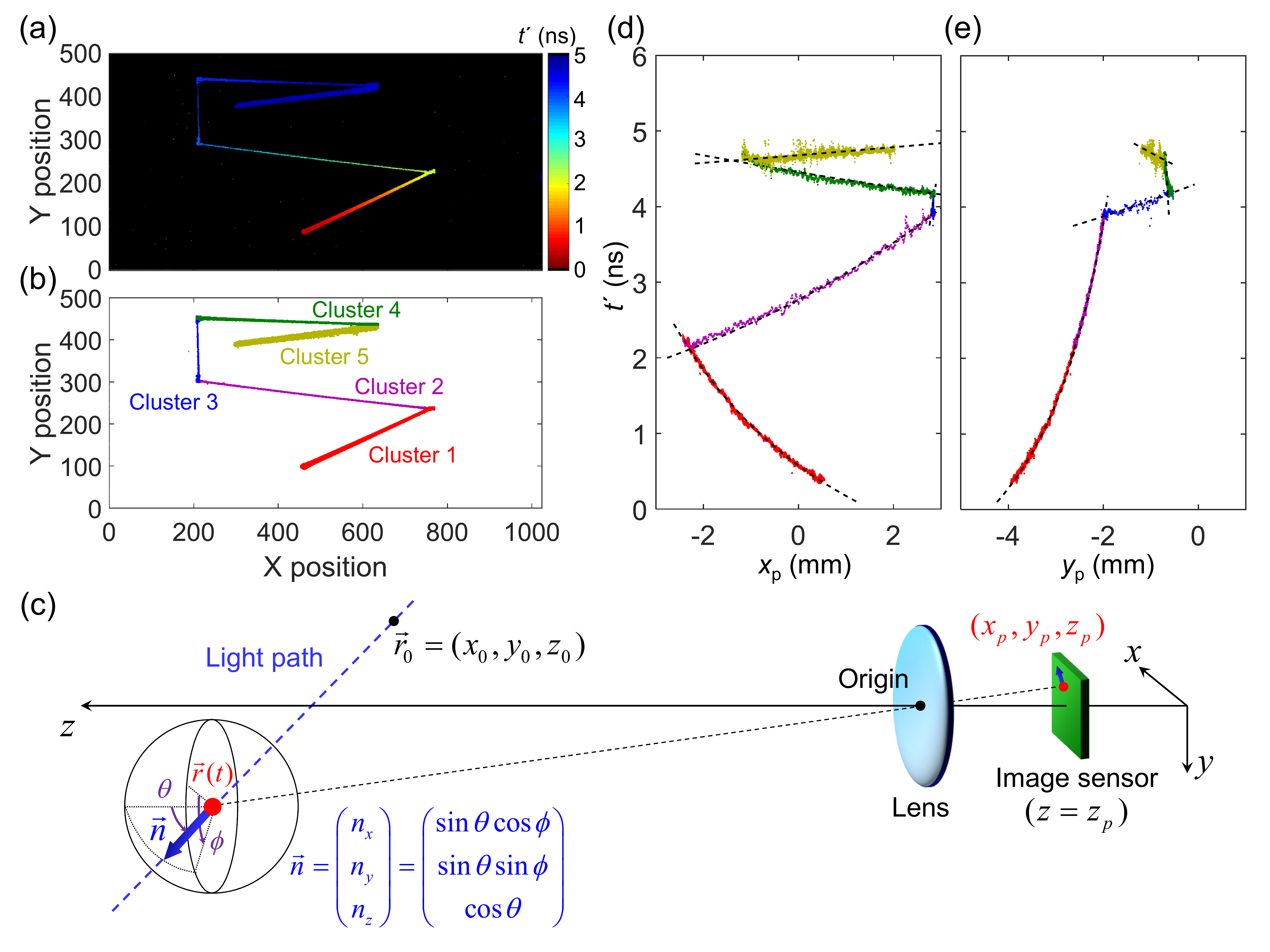}
\caption{\label{analysis}Analysis of extracted spatio-temporal data set. (a) Color-coded plot of observation time for the light path. Red segments are observed earlier, and blue segments are observed later. Dark regions have no timing data. (b) The light path data is subdivided into five data clusters of straight paths using 2D GMM fitting. (c) Schematic view of coordinate system for light-in-flight reconstruction. The laser beam image is projected onto the focal plane by an objective lens in a flipped geometry. (d) Observation time $t^{\prime}$ as a function of $x_p$, the $x$-position on the focal plane. Each data cluster is individually fitted using a theoretical formula. Dashed lines are the fitting curves. (e) $t^{\prime}$ as a function of $y_p$.}
\end{figure*}
Fig.~\ref{video_frames}(a) describes the principle of time gate scanning method. A 3.8 ns global gate window is synchronized to a pulsed laser source. The pixels detect only photons impinging during the gate window. The gate position can be scanned with a gate shift of 36 ps relative to the laser pulse. At each gate position, 255 binary photon-counting frames are summed to form an 8-bit image. 250 slices of 8-bit images with the scanned time-gating window position over 9 ns range are used to extract the timing information of impinging photons. The gate window is defined by sending a short electrical pulse to the gate switch in the pixels. The data acquisition time for full gate scanning (250 slices) is 15 to 30 seconds, much shorter than in previous works (e.g. hours~\cite{Velten,Velten2} to 10 minutes~\cite{Gariepy}). This can be potentially reduced further by introducing a higher power laser.\\
Fig.~\ref{video_frames}(b) shows the procedure to generate a 3-dimensional light-in-flight video. The captured 250 slices of time-gating images form a photon intensity distribution for each pixel as a function of gate position. A rising edge position $\tau^{\prime}$ and “high” level of photon intensity $I$ in the intensity profile can be extracted for each pixel. Time-of-arrival information $t^{\prime}$ over the array is obtained by subtracting independently measured pixel-position-dependent timing skew from $\tau^{\prime}$. The intensity profile is given as a convolution of the arriving photon distribution and the gating profile, and the laser pulse width appears to be elongated by 3.8 ns in the raw intensity profile. Analogously to a previously reported temporal deconvolution technique~\cite{Gariepy}, the laser pulse width can be narrowed to the picosecond scale by replacing the intensity profile with a Gaussian distribution having a mean value of $t^{\prime}$, standard deviation of 150 ps (corresponding to the combined jitter of laser and detector), and integrated photon counts of $I$. Fig.~\ref{video_frames}(c) shows the selected frames of a reconstructed 3-dimensional light-in-flight video, where each frame is superimposed with a 2-dimensional intensity image independently captured by the same SPAD camera. A laser pulse starts to be observed around $t^{\prime}$ = 0 ns, is reflected by multiple mirrors, and goes out of sight around $t^{\prime}$ = 5.832 ns. The laser beam at $t^{\prime}$ = 5.112 ns appears to be stretched compared to the beam at $t^{\prime}$ = 2.232 ns. This implies an enhanced apparent velocity of the beam coming towards the camera at $t^{\prime}$ = 5.112 ns. Note that our system requires no mechanical laser scanning, spatial interpolation and dark noise image subtraction owing to its high spatial resolution array and low dark count rate~\cite{Morimoto}. A full 3-dimensional light-in-flight video with the frame interval of 36 ps is available in Supplementary Video 2.

\section{\label{sec:level4}4-DIMENSIONAL RECONSTRUCTION OF LIGHT-IN-FLIGHT}
\begin{figure*}[tb]
\includegraphics[width=1.8\columnwidth]{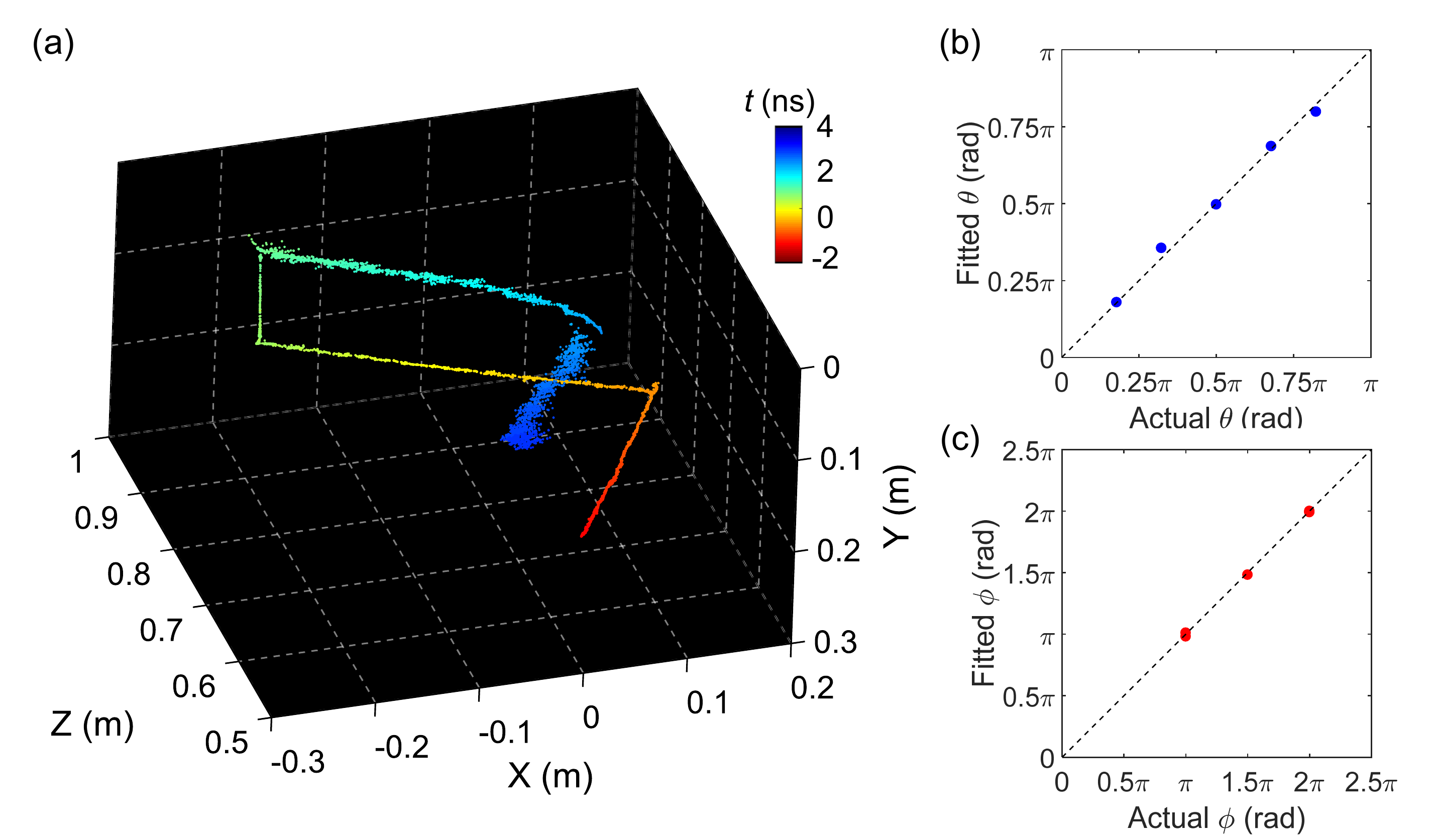}
\caption{\label{4D_map}Reconstructed 4-dimensional light-in-flight observation. (a) 4-dimensional point cloud reproduced from the measured 3-dimensional spatio-temporal data. The origin of the Cartesian coordinate system is defined at the optical center of the lens for the SPAD camera. The ($x$, $y$, $z$, $t$) coordinates of each data point are reproduced using the obtained fitting parameters for the corresponding data cluster. (b) and (c) Comparison of fitted light propagation angles $\theta$ and $\phi$, respectively, with the actual angles for five data clusters. The fitted results show a good agreement with the actual angles. }
\end{figure*}
Fig.~\ref{analysis}(a) shows a color-coded plot of the observation time $t^{\prime}$ for the propagating light. To reconstruct 4-dimensional light path, the data set needs to be subdivided into straight paths. We adopted the 2D Gaussian mixture model (GMM) fitting for data clustering, which is a common unsupervised machine learning technique. Fig.~\ref{analysis}(b) shows the data points separated into five clusters. Note that the data clustering is performed based on 2-dimensional spatial information ($x$ and $y$), whereas the time information $t^{\prime}$ is not taken into account.\\
Fig.~\ref{analysis}(c) depicts the coordinate system for the light-in-flight analysis. Here we derive a theoretical formula to perform the least square regression for estimation of 4-dimensional light trajectory:
\footnotesize
\begin{align}
  \label{7_main2}
   (\vec{r}_0, \theta, \phi) = \argmin_{\vec{r}_0, \theta, \phi} [&\sum^N_{i} \{x_p^i - \frac{z_p}{z_0 + c\cdot f(t^{\prime i})\cdot \cos{\theta}}\cdot \nonumber\\
   &(x_0 + c\cdot f(t^{\prime i})\cdot \sin{\theta}\cos{\phi})\}^2 \nonumber\\
  + &\sum^N_{i} \{y_p^i - \frac{z_p}{z_0 + c\cdot f(t^{\prime i})\cdot \cos{\theta}}\cdot \nonumber\\
  &(y_0 + c\cdot f(t^{\prime i})\cdot \sin{\theta}\sin{\phi})\}^2],
\end{align}
\begin{eqnarray}
  \label{7_ft}
   f(t^{\prime}) = \frac{1}{2}\cdot \frac{c^2 t^{\prime 2} - (x_0^2 + y_0^2 + z_0^2)}{c^2 t^\prime + c(x_0 \sin{\theta}\cos{\phi} + y_0 \sin{\theta} \sin{\phi} + z_0 \cos{\theta})}.
\end{eqnarray}
\normalsize
where $N$ is the total number of measurement data points, ($x_p^i$, $y_p^i$) the corresponding pixel position on the focal plane for $i$-th data point, $-z_p$ the focal length (12.5mm), $t^{\prime i}$ the measured observation time for $i$-th data point, $\vec{r}_0=(x_0, y_0, z_0)$ the position of the beam at $t$ = 0, and the normalized light propagation vector is represented as $\vec{n} = (\sin{\theta}\cos{\phi}, \sin{\theta}\sin{\phi}, \cos{\theta})$ using a polar coordinate system. A single straight light trajectory can be expressed as $\vec{r} = \vec{r_0} + ct\cdot\vec{n}$, defined by five fitting parameters, $x_0$, $y_0$, $z_0$, $\theta$, $\phi$. Those parameters are estimated by solving the above optimization problem for each data cluster.\\
Fig.~\ref{analysis}(d) and~\ref{analysis}(e) show the measured observation time $t^{\prime}$ as a function of $x_p$ and $y_p$, where ($x_p$, $y_p$) is the position of the pixel on the focal plane. Different colors of the data points correspond to the different data clusters. The result of fitting for each data cluster is shown as a dashed line, exhibiting a good agreement with the measured points for both $x_p$ and $y_p$. Note that the fitting involves five independent parameters and is sensitive to the variation of the data. Hence, the spatio-temporal resolution of the camera is a critical factor determining the accuracy of the 4-dimensional reconstruction.\\
As shown in Fig.~\ref{4D_map}(a), the 4-dimensional point cloud is reconstructed without prior knowledge. The time evolution t of the beam is shown as colors of points, having a certain offset from the observation time $t^{\prime}$ due to the finite traveling time from the scattering point to the camera. The fully-recovered 4-dimensional information for each point enables the visualization of the point cloud from arbitrary viewpoint apart from the actual location of the SPAD camera. In contrast to the 3-dimensional light-in-flight video where the apparent velocity of light changes as a function of propagation angle, the 4-dimensional point cloud indicates the uniform speed of light irrespective of the beam position and angle. Relatively larger variation of the data points is observed for cluster 4 and 5. This originates from the enhanced apparent velocity of light coming towards the camera, leading to the reduced fitting accuracy and increased deviation.\\
The accuracy of the reconstruction is evaluated in Fig.~\ref{4D_map}(b) and ~\ref{4D_map}(c). The figures show the fitted light propagation angles $\theta$ and $\phi$ as a function of the actual geometrical angles. Five dots correspond to the fitting results of the five data clusters. All the dots are along with the ideal trend (dashed line), indicating that the fitting results are in good agreement with the actual light propagation vectors.\\
Detailed analysis of the reconstruction performance for various parameter combinations is investigated based on Monte Carlo simulation in the Supplementary Note. The simulation shows that with the horizontal resolution of 1024 pixels, the estimation errors of the fitting parameters improve by a factor of 6 to 9 with respect to previously reported 32$\times$32 pixel SPAD cameras, thereby justifying the necessity of megapixel resolution for high-precision 4-dimensional light-in-flight imaging. The results also suggest that a further increase of the pixel array size could have a non-negligible impact on the stability and accuracy of light-in-flight measurements, along with the potential of further expansion of the measurement in distance and field-of-view.

\section{\label{sec:level5}DISCUSSION}
Our approach of reconstructing extra-dimensional light-in-flight information with high spatio-temporal resolution single-photon camera can be applied to a wide range of high-speed imaging techniques. One of the promising applications is the non-line-of-sight imaging~\cite{Kirmani,Velten3,Katz,OToole2,Liu}; by extending our approach for concentrated linear beam towards a diffused light, hidden objects can potentially be imaged with more simplified setup. The technique could also be used for imaging around and behind obstructing objects. Another potential application is the combination with optical tomography technique~\cite{Huang,Durduran,deHaller}; generalization of our theory can be useful for non-invasive 3-dimensional monitoring inside a target object structure as well as non-destructive measurement of 3-dimensional distribution of physical parameters such as refractive index and transmittance.\\

\appendix
\section*{APPENDIX: DERIVATION OF THEORETICAL FORMULA FOR LEAST SQUARE REGRESSION}
Time evolution of laser pulse position $\vec{r}(t)$ can be described as $\vec{r}(t)=(x(t), y(t), z(t))=\vec{r}_0+ct\cdot\vec{n}$, where $\vec{r}_0=(x_0, y_0, z_0)$ is the time-independent constant vector, $c$ the speed of light in vacuum, and $\vec{n}=(n_x, n_y, n_z)$ the normalized vector representing the direction of light propagation. Note that $t$ is the time when the laser pulse reaches position $\vec{r}(t)$, and has an offset from $t^\prime$, the time when the laser pulse at $\vec{r}(t)$ is observed at the camera position. The laser pulse position projected to the image sensor plane (focal plane) is $\vec{r}_p(t)=(x_p, y_p, z_p)=\alpha(t)\cdot\vec{r}(t)$, where $\alpha(t)$ is the time-dependent coefficient, and $-z_p$ the focal length (12.5 mm). Given that $z_p$ is time-independent, $\alpha(t)$ can be written as $\alpha(t)=z_p/(z_0+ct\cdot n_z)$. The movement of the laser pulse projected to the image sensor is described as:
\footnotesize
\begin{eqnarray}
  \label{7_xp}
   x_p(t) = \frac{z_p}{z_0+ct\cdot n_z}\cdot (x_0+ct\cdot n_x), \nonumber\\ 
   y_p(t)
   = \frac{z_p}{z_0+ct\cdot n_z}\cdot (y_0+ct\cdot n_y).
\end{eqnarray}
\normalsize
Considering the light propagation time from $\vec{r}(t)$ to the camera, the observation time $t^\prime$ can be written as
\footnotesize
\begin{eqnarray}
  \label{7_tprime}
   t^\prime = t + \frac{|\vec{r}(t)|}{c} = t + \frac{1}{c}\cdot \sqrt{|\vec{r}_0|^2 + 2ct(\vec{r}_0\cdot \vec{n}) + c^2 t^2}.
\end{eqnarray}
\normalsize
Solving this equation will give the following:
\footnotesize
\begin{eqnarray}
  \label{7_t}
   t = f(t^\prime) = \frac{1}{2}\cdot \frac{c^2 t^{\prime 2} - |\vec{r}_0|^2}{c^2 t^\prime + c(\vec{r}_0\cdot \vec{n})}.
\end{eqnarray}
\normalsize
By substituting Eq. (\ref{7_t}) to Eq. (\ref{7_xp}), the projected laser pulse position as a function of the observation time $t^\prime$ is expressed as:
\footnotesize
\begin{eqnarray}
  \label{7_xp2}
   x_p(t^\prime) = \frac{z_p}{z_0 + c\cdot f(t^\prime)\cdot n_z}\cdot (x_0 + c\cdot f(t^\prime)\cdot n_x), \nonumber\\
   y_p(t^\prime) = \frac{z_p}{z_0 + c\cdot f(t^\prime)\cdot n_z}\cdot (y_0 + c\cdot f(t^\prime)\cdot n_y).
\end{eqnarray}
\normalsize
From the time-resolved measurement, $N$ sets of 3-dimensional data points $(x^i_p, y^i_p, t^{\prime i}$) are obtained ($i$ = 1,2, …, N). To reconstruct 4-dimensional light-in-flight history, the six parameters $x_0$, $y_0$, $z_0$, $n_x$, $n_y$, $n_z$ have to be determined by solving the following optimization problem:
\footnotesize
\begin{align}
  \label{7_main}
  (\vec{r}_0, \vec{n}) = \argmin_{\vec{r}_0, \vec{n}} [&\sum^N_{i} \{x_p^i - \frac{z_p}{z_0 + c\cdot f(t^{\prime i})\cdot n_z}\cdot \nonumber\\
  &(x_0 + c\cdot f(t^{\prime i})\cdot n_x)\}^2 \nonumber\\
  + &\sum^N_{i} \{y_p^i - \frac{z_p}{z_0 + c\cdot f(t^{\prime i})\cdot n_z}\cdot \nonumber\\
  &(y_0 + c\cdot f(t^{\prime i})\cdot n_y)\}^2],
\end{align}
\normalsize
The Eq. (\ref{7_main2}), (\ref{7_ft}) can be derived by converting Eqs. (\ref{7_main}) and (\ref{7_t}), respectively, to a polar coordinate system, where $\vec{n} = (\sin{\theta}\cos{\phi}, \sin{\theta}\sin{\phi}, \cos{\theta})$.\\
Note that data cluster comprising of smaller number of data points could have a convergence problem during the least square regression. This can be avoided by taking a continuity assumption for the light path into consideration; starting point of the target data cluster should coincide with the end point of the previous data cluster. In practice, this assumption can be implemented by adding an extra cost function $\lambda \cdot \{(x_c-x_0-ct_c\cdot n_x)^2+(y_c-y_0-ct_c\cdot n_y)^2+(zc-z_0-ct_c\cdot n_z)^2\}$ to Eq. (\ref{7_main}), where ($x_c$, $y_c$, $z_c$, $t_c$) is the 4-dimensional coordinate for the end point of the previous data cluster, and $\lambda$ the positive coefficient.

\begin{acknowledgments}
We thank C. Bruschini and S. Frasca for stimulating discussions. This research was funded, in part, by the Swiss National Science Foundation Grant 166289 and by Canon Inc.
\end{acknowledgments}


\nocite{*}

\bibliography{apssamp}

\end{document}